# Computational Models for Attitude and Actions Prediction

Jalal Mahmud[1], Geli Fei[2*], Anbang Xu[1], Aditya Pal[3*], Michelle Zhou[4*]

[1]IBM Research - Almaden, jumahmud@us.ibm.com, anbangxu@us.ibm.com
[2]University of Illinois at Chicago, gfei2@uic.edu
[3]Facebook Inc, aditya.pal@gmail.com
[4]Juji Inc, mzhou@acm.org

**Abstract**

In this paper, we present computational models to predict Twitter users' attitude towards a specific brand through their personal and social characteristics. We also predict their likelihood to take different actions based on their attitudes. In order to operationalize our research on users' attitude and actions, we collected ground-truth data through surveys of Twitter users. We have conducted experiments using two real world datasets to validate the effectiveness of our attitude and action prediction framework. Finally, we show how our models can be integrated with a visual analytics system for customer intervention.

## Introduction

In the last few years, social media such as Twitter has emerged, and different brands have social media presence to attract their potential customers. People express various opinions about such brands in social media. Some people may like a brand (e.g., Delta Airlines, Fitbit), some may show neutral attitude, and others may dislike the brand. Some may have formed an attitude towards a brand very recently, and others may have an attitude for quite a long time. Some people have attitude with higher confidence than others, some may remember their attitude well and some are more likely to change their attitudes. People also take different actions (e.g., buy a product/service corresponding to that brand, recommend others to buy) based on their attitude about a brand.

As a concrete example, consider a Twitter user John who has somewhat negative attitude towards Delta Airlines. He just started to use this airline, remembers his attitude well, not very confident about his current attitude and likely to change his attitude if Delta Airlines offers a better service. He is also somewhat likely to buy a ticket of his next trip from Delta Airlines, but is unlikely to recommend others to use this airline. Another user Zen has very positive attitude towards Delta Airlines, using this for many years, remembers her attitude well, is quite confident about her attitude, and is unlikely to change her attitude even if quality of service offered by Delta slightly drops. She is very likely to buy a ticket from Delta Airlines for her next trip, and recommend her friends to use this airline Prior works on sentiment/opinion analysis (Li et al. 2010, Li et al. 2012) can be useful to know whether a user may like/dislike a brand. A recent work on attitude modeling (Gao et al. 2014) also describe inferring attitude towards controversial topics in terms of sentiment, opinion and actions. However, such works do not address whether such attitudes are persistent (e.g., an individual formed an attitude for a long time) or temporary (e.g., an individual formed an attitude recently). They also do not provide the strength of attitude (e.g., whether the individual has attitude with high/low confidence). Furthermore, they do not address how well a user remembers his/her attitude or whether the user is likely to change the attitude. Such fine-grained information of consumer attitude can be useful for social media marketers who would directly engage such consumers on social media platforms (Jansen et al. 2009). Furthermore, predicting social media actions (e.g., retweeting a tweet) as described in Gao et al. (2014) is inadequate for such direct engagement scenario where marketers would be more interested to know whether a consumer will take actions outside the social media (e.g., buying a product of that brand, or recommending a friend).

Motivated by such a need, we present computational models to predict a Twitter user's attitude in terms of a number of characteristics such as *attitude favorability* (How much a consumer likes or dislikes an attitude object), *attitude persistence* (whether an attitude is persistent), *attitude confidence* (strength of attitude), *attitude accessibility* (How well a consumer remembers attitude about the object) and *attitude resistance* (How likely a consumer keeps the present attitude). Our work is inspired by marketing literature where attitude is described in terms of such characteristics (Hoyer et al. 2008).

Since there are no publicly available ground truth data of attitude characteristics, we have collected such data using self-report surveys conducted among Twitter users. Using the ground-truth data, we developed statistical models to predict users' attitude. Our classification based models of attitude characteristics are based on features extracted from users' historical tweets. Such models can classify whether a

---

user has specific characteristics of attitude, and also output the likelihood of those attitude characteristics. We have also developed statistical models to infer likelihood of different action intentions[1] (e.g., intention to buy a product) based on one's attitude. Similar to our models for predicting different attitude characteristics, our models for predicting action intentions are also trained from features derived from users' historical tweets. Since a person may not have any attitude towards a particular brand, it is useful to first identify potential Twitter users who have attitude towards the brand and then apply attitude and action models. In this work, we assume that users who mentioned a brand has an attitude towards the brand, and use simple keyword filtering to identify such users before applying attitude and action models.

We performed extensive experiments using two real world datasets to validate the effectiveness of our models. For attitude characteristics, we observed mixed result. While attitude favorability can be predicted within 65-69% AUC, prediction accuracies of other characteristics are low to moderate (52-59% AUC). Action intentions can be predicted within 56-67% AUC. We have also integrated our prediction models with a visual analytics system that recommends Twitter users with specific attitude towards a brand, and thus allows potential intervention. Below we list the summary of contributions of this work:

- A survey study of understanding attitude of Twitter user towards multiple brands.
- Models to predict attitude of Twitter users towards a brand in terms of a set of characteristics.
- Models to predict users' intention to take different actions based on their attitudes.
- Experiments demonstrating the effectiveness of our models.
- A visual analytic system that integrates our models and allows customer intervention.

## Related Work

Our work is related to a number of prior works on attitude modeling, sentiment/opinion analysis, and behavior prediction.

The work most closely relates to ours is the recent work of Gao et. al. on modeling user's attitude towards controversial topics (Gao et al. 2014). Their work is inspired by a theoretical framework in psychological and marketing research on attitudes and attitude model, where attitude is defined as a unified concept containing three aspects: "feelings", "beliefs", and "actions" (McGuire et al. 1968, Eagly et al. 1993, Schiman et al. 2010).

Their model can capture such aspects, and can predict user's attitude in terms of them. However, they have trained models from ground-truth observable from Twitter data that has a number of limitations. First, in their model re-tweets are considered as an action towards supporting a topic, which may not be reliable when re-tweets may not mean an endorsement. Second, observable ground truth of Twitter actions (such as re-tweets, tweeting) cannot capture actions (e.g., buying a product or recommending others to buy) or action intentions (intention to buy a product) which has impact beyond social media. In contrast to Gao et al., we have adopted a survey-based approach to collect ground-truth of attitudes and action intentions from Twitter users. This gives us reliable ground-truth data as well as provides us an opportunity to model action intentions that can impact beyond social media (e.g., intention to buy a product). Furthermore, we have also modeled fine-grained attitude characteristics such as attitude persistence, attitude confidence, attitude accessibility and attitude resistance, which were beyond the scope of Gao et al (2014).

Our model of attitude favorability is also related to a wide number of researches on opinion and sentiment analysis (Pang et al. 2008, Abu-Jbara et al. 2012, Li et al. 2012). There are previous works to detect sentiment from various forms of text data such as documents, blogs, and tweets. Li et al. described a Topic-Level Opinion Influence Model (TOIM) that jointly incorporates topic factor and social influence in a two-stage probabilistic framework (Li et al. 2012). Lin et al. proposed an unsupervised probabilistic modeling framework based on Latent Dirichlet Allocation (LDA) to detect sentiment and topic simultaneously from text (Lin et al. 2009). Li et el. described sentiment detection from micro-blogs using collaborative learning approach (Li et al. 2010). Recently, Hu et al. proposed sentiment analysis using social signals in both supervised and unsupervised ways (Hu et al. 2013a, Hu et al. 2013b). Prior works also described aggregating message level sentiments to infer a social media user's overall sentiment toward a target, or predicting a user's opinion/sentiment toward a topic/target. Kim et al. proposed user-level sentiment prediction using collaborative filtering approach (Kim et al. 2013). Tan et al. described user-level sentiment analysis using social network information (Tan et al. 2011). We are inspired by existing works on user level sentiment analysis towards a topic/target to develop supervised models of attitude favorability. However, in this work we also investigate how this attitude component affects prediction of other attitude components in a joint prediction using iterative classification.

Our prediction of action intentions (e.g., buy/recommend) from social media updates of users relate to a wide number of prior works on behavior prediction from social media (Mahmud et al. 2013, Feng et al. 2013, Yang et al. 2010, Hannon et al. 2010). Such works predict observable social media action/behavior such as replies, retweets or follow behavior. In contrast, we aim to predict action intentions, which are not observable in social media. Our models to predict such action intention is also developed from ground truth collected from self-reported action intentions

---

[1] In this paper, we use the term action, action intention and intention interchangeably.

in-contrast to observable ground truth such as replies, re-tweets.

| Attitude Characteristics | Questions |
|---|---|
| Favorability | How much have you liked your travel experience with Delta Airlines? |
| Persistence | How often have you used Delta Airlines for your travel? How long have you used Delta Airlines for your travel? |
| Confidence | Based on your answers, how certain are you about your answers? |
| Accessibility | How well do you remember your attitude about Delta Airlines? |
| Resistance | How likely will you switch to another airlines if Delta reduces efficiency of service? How likely will you switch to another airlines if Delta reduces comfort of service? How likely will you switch to another airlines if Delta increases cost of service? |

**Table 1. Questions to assess attitude about Delta**

| Action Intentions | Questions |
|---|---|
| Buy | How likely are you going to buy ticket of your next trip from Delta airlines? |
| Recommend | How likely are you going to recommend others to fly by Delta airlines? |
| Prohibit | How likely are you going to tell others not to fly by Delta airlines? |

**Table 2. Questions in Delta survey to assess action intentions**

## Survey Study - Methodology

Since there is no publicly available dataset on attitude of Twitter users towards a product/service, we collected two real-world datasets from Twitter. Our first dataset is about "Delta" airlines, and the second dataset is about "Fitbit" exercise equipment. Note that, we wanted to collect datasets for two popular brands, and at the same time brands which are quite different. Both the brands are quite popular (lot of people talk about them in Twitter) and they represent two different market segment: travel and health-care.

**Survey Questions:** We created the surveys using SurveyGizmo (http://www.surveygizmo.com/), a popular survey-building tool. Our survey questions were designed to capture various aspects of attitude (Hoyet et al. 2008). Table 1 and 2 shows questions that were included in the Delta survey. Similarly, Table 3 and 4 shows questions that were included in the Fitbit survey. Response to each question was on 5-point Likert scale. Note that, some attitude variables are measured using multiple questions. Response to such variables is computed as an average of the response of those questions. A high value of favorability means user likes the brand more, and low value means the opposite. A high/low value of persistence means user has more/less persistence attitude. Similarly, a high value of accessibility means user can remember the attitude easily, and low value means the opposite. A high value of resistance means user is more likely to stay with the brand, and low value means the opposite. A low value on the response to each action intention means the user is less likely to perform the action, and high value means the opposite. In both surveys, we also asked each participant what other actions they want to take based on their attitude so as to gain some insights.

| Attitude Characteristics | Questions |
|---|---|
| Favorability | How much have you liked your experience with Fitbit? |
| Persistence | How long have you used Fitbit? |
| Confidence | Based on your answers, how certain are you about your answers? |
| Accessibility | How well do you remember your attitude about Fitbit? |
| Resistance | How likely will you switch to another fitness device if Fitbit reduces efficiency (e.g., calorie tracking)? How likely will you switch to another fitness device if Fitbit reduces comfort (e.g., comfort to wear)? How likely will you switch to another fitness device if Fitbit increases cost? How likely will you switch to another fitness device if Fitbit reduces visual attractiveness? |

**Table 3. Questions to assess attitude about Fitbit**

| Action Intentions | Questions |
|---|---|
| Buy | How likely are you going to buy next fitness device from Fitbit? |
| Recommend | How likely are you going to recommend others to buy fitness device from Fitbit? |
| Prohibit | How likely are you going to tell others not to use Fitbit? |

**Table 4. Questions in Fitbit survey to assess action intentions**

**Survey Participants:**
We identified 7534 Twitter users who tweeted about "Delta" airlines during March 2014 to June 2014. They were identified using the keyword @delta in Twitter's Search API. We sent them requests to participate in our survey. To do so, we constructed 6 Twitter accounts and these accounts were used for sending the surveys. We wanted to create multiple Twitter accounts to engage Twitter users in our survey so that we could overcome rate limits associated with sending many messages from a single Twitter account and also risk of being suspended for sending too many similar messages. These accounts were constructed in a way to appear as genuine as possible, so that a survey request from them does not appear to be a phishing or marketing campaign to target users and the risk of them getting marked as spam by Twitter is mitigated. There was no specific reason for creating 6, however, that many accounts were adequate for our message sending purpose.

|  | Mean | Std. Dev. | Correlations | | | | | | |
|---|---|---|---|---|---|---|---|---|---|
|  |  |  | 1 | 2 | 3 | 4 | 5 | 6 | 7 |
| 1. Favorability | 3.15 | 1.34 |  |  |  |  |  |  |  |
| 2. Persistence | 3.23 | 0.93 | **.43** |  |  |  |  |  |  |
| 3. Confidence | 3.68 | 0.94 | **.41** | **.45** |  |  |  |  |  |
| 4. Accessibility | 4.21 | 0.67 | .03 | -0.07 | -.016 |  |  |  |  |
| 5. Resistance | 2.86 | 0.89 | **.32** | **.17** | **.13** | -0.02 |  |  |  |
| 6. Buy | 3.09 | 1.24 | **.77** | 0.44 | .37 | .042 | **.26** |  |  |
| 7. Recommend | 2.9 | 1.23 | **.81** | **.35** | **.35** | .059 | **.28** | **.84** |  |
| 8. Prohibit | 2.29 | 1.19 | **-0.50** | **-0.20** | **-0.13** | -0.01 | **-0.15** | **-0.43** | **-0.45** |

**Table 5. Descriptive Statistics of Variables from the Delta Survey Study** (Significant correlations are shown in bold).

|  | Mean | Std. Dev. | Correlations | | | | | | |
|---|---|---|---|---|---|---|---|---|---|
|  |  |  | 1 | 2 | 3 | 4 | 5 | 6 | 7 |
| 1. Favorability | 4.0 | 1.12 |  |  |  |  |  |  |  |
| 2. Persistence | 3.0 | 1.44 | **.41** |  |  |  |  |  |  |
| 3. Confidence | 3.89 | 0.99 | **.48** | **.55** |  |  |  |  |  |
| 4. Accessibility | 3.82 | 0.91 | .066 | -0.06 | .046 |  |  |  |  |
| 5. Resistance | 2.9 | 0.93 | **.42** | .16 | **0.244** | -0.078 |  |  |  |
| 6. Buy | 3.5 | 1.11 | **.66** | **.49** | **0.516** | -0.116 | **.402** |  |  |
| 7. Recommend | 3.7 | 1.17 | **.60** | **.44** | **0.536** | -0.033 | **.456** | **.92** |  |
| 8. Prohibit | 2.14 | 1.19 | -.173 | -0.15 | -0.159 | 0.11 | -0.013 | -0.099 | -0.174 |

**Table 6. Descriptive Statistics of Variables from the Fitbit Survey Study** (Significant correlations are shown in bold).

We offered $50 Amazon gift card to 1 out of every 100 survey participants. Similarly, we identified 5261 Twitter users who tweeted about "Fitbit" exercise equipment (identified using the keyword @fitbit in Twitter's Search API), and sent request to participate in our survey. Similar to our "Delta" airlines survey, we created 6 Twitter accounts to send survey requests to Twitter users and we offered $50 Amazon gift card to 1 out of every 100 survey participants. We ensured that each participant took the survey only once.

**Survey Responses:**
823 users responded to our Delta survey (10.9% response rate) and 507 users responded to our Fitbit (9.6% response rate) survey. We manually inspected survey responses and removed incomplete responses. Finally, we had 751 survey responses for Delta and 447 survey responses for Fitbit. For each user who responded, we collected their most recent 3200 tweets (max limit enforced by Twitter) using Twitter's REST API. If they had less than 3200 tweets, then we collected all their tweets. We computed mean, standard deviation and correlations between attitude variables and actions. Table 5 and 6 shows such statistics for Delta and Fitbit dataset.

*What Other Actions People Intend to Take?*
Some respondents mentioned that they would like to share their experience in social media (*"I will share in Facebook my travel experience about Delta", "share how wonderful fitbit is", "Taking to Twitter. Letting my followers know.", "Tweet a complaint, survey, write email complaint, blog a complaint, write to a popular travel blog"*). Some others wanted to share with people outside of social media (*"tell my friends about delta", "I tell everyone I know in real life and interact with on social media how terrible Delta is."*). Some people mentioned that they want to complain to customer service (*"I will report complain to their customer service department next time", "sending a complaint to fitbit"*). Other responses include telling others to cancel membership (*"Don't plan to buy ticket from Delta, I will tell my friends to cancel their membership"*), increasing status ("*I will try to get the silver status"),* participate in more surveys (*"I will participate in surveys to tell my feedback about Delta"*).

## Classification Approaches

Here we describe classification models to classify various attitude dimensions and action intentions.

**Attitude Characteristics Classification**
We developed statistical models to classify each attitude characteristics. Such models used a set of features extracted from users' historical tweets. For simplicity, we developed binary classifiers for each attitude characteristics. Thus, we first converted each attitude characteristics into binary values (using mean of scale as threshold).

*Feature Extraction.*
We extract a set of features for attitude classification, which are described below.

- *Unigram features:* This feature represents all unigrams extracted from tweets.
- *Sentiment features*: We use a sentiment/opinion dictionary[2] that contains a list of words with their positive/negative sentiment polarity. We count total number of positive/negative words in user's tweets and used that positive and negative counts as features.
- *Context-based Sentiment/Opinion feature*: This is similar to the above sentiment/opinion feature; however, it looks for sentiment words that appear in the surrounding area of the brand name (e.g., textual patterns like "awesome delta"). Thus, we counted how many times positive sentiment words in the dictionary co-occur with the brand name and how many times negative sentiment words in the dictionary co-occur with the brand name and used those counts as feature values.
- *Domain-specific sentiment feature*: This feature is similar to the above *Context-based Sentiment/Opinion feature*, however, it is computed by matching words in users' tweets with a domain-specific sentiment dictionary. Words in such dictionary does not appear in general sentiment dictionary. Domain-specific sentiment dictionary is constructed from training users' tweets. From such tweets, we compute a sentiment score for each word that co-occurs with the brand keyword (e.g., delta) where such score represents how likely this word appears with the brand word in positive/negative context. If positive score for a word outperforms negative score by a certain threshold or vice versa, such word is added to the domain specific sentiment dictionary.
- Length of use feature: This feature aims to capture the attitude-persistence of a user. It is obtained by taking the timestamp difference of a user's latest and oldest mention of the brand.
- *Frequency feature*: This feature represents how often the user mentions the brand.

*Statistical Models.*
Once we computed the above mentioned features, we developed statistical models using WEKA (Witten et al. 2011). We tried a number of classifiers such as Naive Bayes, SMO (SVM), Random Forest from WEKA and performed 5-fold cross validation. SMO and Random Forest based classifier achieved comparable performance. In experiment section, we report experimental result for SMO classifier.

**Action Classification**
Similar to our approach for attitude classification, we developed a set of binary classifiers for classifying each action intention. Thus, we converted each action into binary values (using mean of scale as threshold). Classifiers for action classification used similar features as described for attitude classification. We tried state of the art classifiers such as Naive Bayes, SMO (SVM), Random Forest from WEKA, and performed 5-fold cross validation. We found that comparable performance from SMO and Random Forest. In experiment section, we report experimental result for SMO classifier.

**Iterative Classification Approach**
As we have shown in the former section, strong correlations can be found among different attitude characteristics and actions. Naturally, we ask the question if we could improve our prediction by leveraging the correlations among these variables. In this paper, we propose to use Iterative Classification Algorithm (ICA) (Sen et al. 2008), one of the most popular approximate inference algorithms for collective classification, to help predict multiple attitude characteristics and actions jointly.

In traditional machine learning community, classification is typically done on each object independently, without taking into account the underlying connections among objects. Collective classification algorithms, in contrast, refer to the combined classification of a set of interlinked objects. In general, exact inference for collective classification is known to be a NP-hard problem, so most of the research in collective classification has been devoted to the development of approximate inference algorithms. ICA is an approach based on local conditional classifiers. ICA assumes we have a classifier $f$ that takes the value of neighboring nodes and returns the best value for current node, which gives us extreme flexibility in choosing $f$.

Since different attitude characteristics and actions are interlinked, they can be considered forming a graph. Each node represents a response variable whose class we need to predict, and an edge in the graph represents two response variables having correlation stronger than a threshold. Each node has two sets of features associated with it: static features and dynamic/relational features. Static features are those which remain unchanged during classification process in testing, while dynamic features are affected as classification process proceeds, i.e. the values of these features change during the testing phase because they reflect the predicted labels of a node's neighbors. In our case, the static features for each attitude characteristics or action are those we introduced for independent prediction; while the dynamic features are the prediction of all neighbor nodes. Although we are interested in predicting binary class labels for each response variable, instead of using either positive or negative label, we use the probability of positive class as the feature value for fine-grained prediction.

---
[2] http://www.cs.uic.edu/~liub/FBS/sentiment-analysis.html#lexicon

1. **for** each $Y_i \in \mathcal{Y}$ **do** // bootstrapping
2.   compute $\vec{a}_i^s$ based on only static features
3.   $y_i \leftarrow f_i^s(\vec{a}_i^s)$
4. **end for**
5. **repeat** // iterative classification
6.   **for** each $Y_i \in \mathcal{Y}$ **do** // compute new estimate of $y_i$
7.     compute $\vec{a}_i^{s+d}$ using static features and current assignment to $\mathcal{N}_i$ as dynamic features
8.     $y_i \leftarrow f_i^{s+d}(\vec{a}_i^{s+d})$
9.   **end for**
10. **until** all class labels have stabilized or after threshold number of iterations

**Figure 1   ICA Algorithm**

The training process works just like traditional supervised learning. However, we need to train two classifiers for each response variable (i.e. attitude characteristics or actions) we want to predict. One classifier is trained using only static features, which is used in the initialization step in testing, because we know nothing about any of the response variables in the beginning; the other classifier is trained using both static features and relational features, which is used in the iterative classification step.

Formally, let $Y_i \in \mathcal{Y}$ be an attitude characteristics or action whose class we want to predict, and $\mathcal{N}_i$ be all neighbor nodes of $Y_i$. For each $Y_i \in \mathcal{Y}$, train two classifiers $f_i^s$ and $f_i^{s+d}$ using $\vec{a}_i^s$ (feature vector for $Y_i$ using only static features) and $\vec{a}_i^{s+d}$ (feature vector for $Y_i$ using both static and dynamic features). The algorithm for testing is shown in Figure 1.

|  | precision | recall | F1 | ROC Area (AUC) |
|---|---|---|---|---|
| Favorability | 0.73 | 0.69 | 0.71 | 0.69 |
| Persistence | 0.59 | 0.62 | 0.61 | 0.59 |
| Confidence | 0.57 | 0.59 | 0.58 | 0.56 |
| Accessibility | 0.56 | 0.54 | 0.55 | 0.53 |
| Resistance | 0.53 | 0.54 | 0.53 | 0.52 |

**Table 7. Result of Attitude Classification (Delta)**

|  | precision | recall | F1 | ROC Area (AUC) |
|---|---|---|---|---|
| Favorability | 0.7 | 0.66 | 0.68 | 0.65 |
| Persistence | 0.58 | 0.57 | 0.57 | 0.57 |
| Confidence | 0.55 | 0.58 | 0.56 | 0.55 |
| Accessibility | 0.54 | 0.53 | 0.53 | 0.52 |
| Resistance | 0.54 | 0.52 | 0.53 | 0.52 |

**Table 8. Result of Attitude Classification (Fitbit)**

In iterative inference methods, proper initialization not only helps an algorithm converge faster, but also helps to achieve better results. For step 3 of Figure 1, we proposed an alternative initialization based on some heuristics to help better initialize the initial states of each node, which in return, gives better results constantly comparing to only using classifier $f^s$. The issue of $f^s$ is that it performs too poorly in predicting some response variables. We only trust $f^s$ when the prediction is confident, meaning the probability of positive class is either higher or lower than a threshold, and we use 0.8 and 0.2 as thresholds in our experiment. While in other cases, we use the prediction of $f^s$ on attitude-favorability to initialize, as this characteristics has the highest accuracy. In such cases, if an attitude characteristics or action has positive correlation with attitude-favorability, we use the predicted value of attitude-favorability to initialize those attitude characteristics or action; otherwise we use one minus the predicted value of attitude-favorability to initialize those attitude characteristics or action. Since using the heuristics gives better results in our experiments, we only report the results based on this approach.

|          | precision | recall | F1   | ROC Area (AUC) |
|----------|-----------|--------|------|----------------|
| Buy      | 0.67      | 0.69   | 0.68 | 0.67           |
| Recommend| 0.66      | 0.65   | 0.65 | 0.64           |
| Prohibit | 0.59      | 0.61   | 0.60 | 0.58           |

**Table 9. Result of Action Classification (Delta)**

|          | precision | recall | F1   | ROC Area (AUC) |
|----------|-----------|--------|------|----------------|
| Buy      | 0.65      | 0.64   | 0.65 | 0.63           |
| Recommend| 0.61      | 0.58   | 0.60 | 0.58           |
| Prohibit | 0.57      | 0.58   | 0.58 | 0.56           |

**Table 10. Result of Action Classification (Fitbit)**

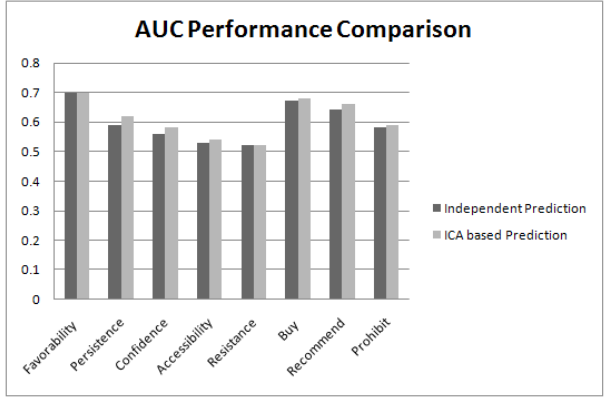

**Figure 2. AUC Performance Comparison between Independent and ICA based prediction (Delta Dataset)**

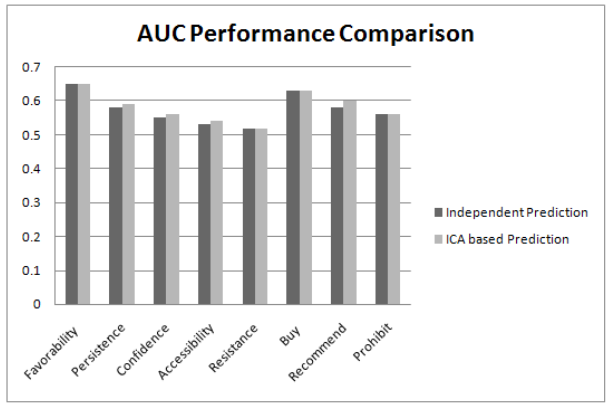

**Figure 3. AUC Performance Comparison between Independent and ICA based prediction (Fitbit Dataset)**

# Experiments

Here we describe experiments we have performed to validate the effectiveness of our approach.

### Attitude Classification
Table 7 and 8 show the result of our attitude prediction for Delta and Fitbit datasets in terms of precision, recall, F1 and AUC. We observe mixed result for attitude characteristics prediction. Attitude favorability can be predicted with reasonable accuracy (more than 65% AUC). However for other attitude characteristics, we found moderate to low prediction accuracy. This is perhaps due to the fact that our survey responders provided more reliable response for attitude favorability questions in comparison to the response of other attitude characteristics (e.g., it might be easier to provide a response on favorability in comparison to resistance). Furthermore, it is more intuitive that textual and sentiment based feature extracted from users' historical tweets contain predictive information to predict attitude favorability (i.e., how much they like/dislike a brand). Attitude accessibility and resistance were more difficult to predict which could be either due to the reason that our survey users did not provide very reliable response for those dimensions or users' historical tweets did not contain enough predictive feature for predicting them.

### Action Classification
Table 9 and 10 show the result of action prediction for both datasets in terms of precision, recall, F1 and AUC. Overall, actions can be predicted within 56% to 67% AUC. In each case, buy action achieves best performance, while prohibit action seems harder to predict. We think one of the difficulties of predicting prohibit action comes from the particularly unbalanced training data, as prohibit action is an extreme action that not many users would do.

### Iterative Classification
Figure 2 and 3 show the comparison of independent prediction and ICA based prediction for both datasets. We see a slight improvement of performance for few attitude characteristics and actions as a result of ICA based prediction. One of the most important reasons that this approach doesn't achieve as good results as we expected is because ideally we hope most classifiers using only static features could achieve reasonably good accuracy. However, due to the difficulty of the problem, only few attitude characteristics and actions could be predicted with reasonable accuracy. By interlinking different dimensions (attitude characteristics and actions), those dimensions with poor prediction bring more noise to the algorithm and even more confuse the classifiers on other dimensions.

# System for Customer Intervention

We integrated our prediction models with a visual analytics system for customer intervention. Such a system is designed to help customer service agents understand customer attitudes towards a brand and support agents to find customers to be targeted for direct intervention (Chen et al. 2013). First, the system uses simple keyword filtering (e.g., @delta, @fitbit) to identify a set of Twitter users who have recently discussed the brand in their tweets. Their attitudes and action intentions are computed based on their tweets and presented to agents for intervention.

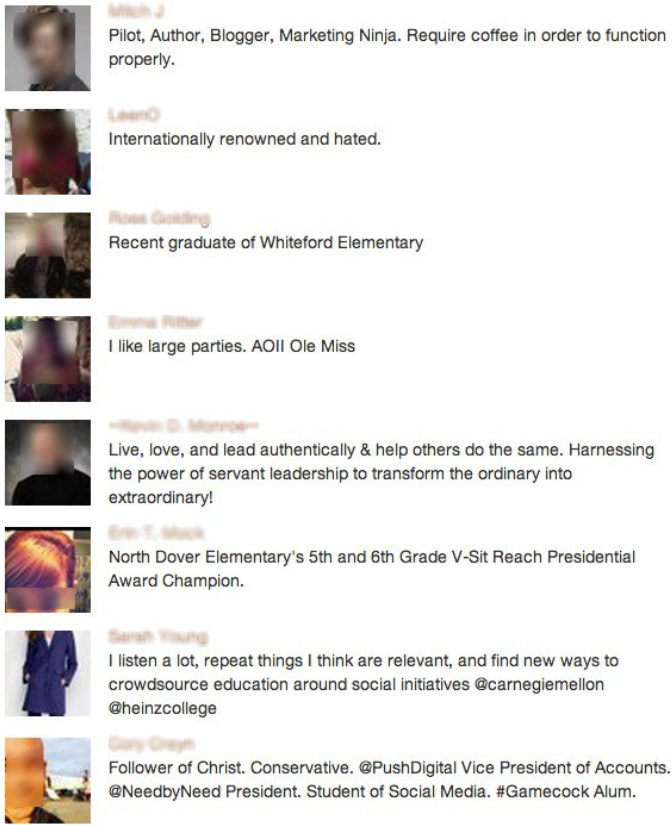

**Figure 4.** A screenshot of the visual analytics system. The system uses simple keyword filtering to identify a set of Twitter users who have recently discussed the brand (delta for this example) in their tweets. The *attitude & intention component* (right) shows the distributions of such customer attitudes and action intentions towards a brand. Individual customers' profiles are shown in the *detailed view component* (left). An agent tries to identify customers who have positive and slightly persistent attitude towards a brand, and are likely to buy its product. She sets visual filters on the *favorability*, *persistence* and *buy* bar charts in the *attitude & intention component*, respectively. Customers within the selected range are shown in the *detailed view component*.

The design is shown in Figure 4 and includes two main components: the *attitude & intention component* and the *detailed view component*. The *attitude & intention component* (right in Figure 4) shows an overview of customer attitudes and action intentions in a number of dimensions including favorability, persistence, confidence, accessibility, resistance, buy, recommend, and prohibit. A bar chart visualizes the distribution of the customers' attitude or action intention values along a dimension. The number in a bar denotes the total number of customers in the corresponding segment (see Figure 4). This overview feature is useful for gaining a holistic understanding of customer attitudes and intentions. The system allows agents to create visual filters on any individual bar chart in the *attitude & intention component* by selecting a range on the axis, and customers within the selected range would then be shown in the *detailed view component* (left in Figure 4) to reflect the selection. These visual filters therefore support selecting target customers according to arbitrary criteria on any of the eight dimensions we provided. The filtering feature helps agents identify target customers for intervention.

The *detailed view component* (left in Figure 4) shows target customers' profiles extracted from social media. Agents can click on a customer's profile to see more detailed information (e.g., the customer's age, location, relevant tweets). This component is designed to provide sufficient contextual information for customer intervention.

## Conclusion

In this paper, we presented models to predict a Twitter user's attitude towards a brand in terms of a set of characteristics, and likelihood to take different actions based on attitudes. Our models are trained and tested using two real-world datasets. We have found that most of the attitude characteristics and action intentions can be predicted with moderate accuracy (60-65%). Furthermore, we have integrated our prediction models to a visualization interface to demonstrate usage in customer intervention. We identify several avenues of future research. First, we want to improve the accuracy of our attitude and action intention prediction models by further exploring available features, and their relationships. For example, we plan to investigate whether users' personality computed from text can help predict attitude towards different brands. Second, we plan to continue our investigation of joint modeling effort to predict attitude and action intentions together. Third, we are interested to explore how to construct attitude models which are easily scalable across multiple brands. For example, it will be interesting to know whether attitude toward one brand can help us learn attitude toward a related brand. Finally, we want to apply our models in real-world intervention scenarios and study their usage.